\documentclass[12 pt,cimento]{article}
\usepackage{epsfig}
\usepackage{changebar}
\usepackage{amssymb}

\begin{document}

\title{The $\Delta$-excitation in the nuclear charge longitudinal response}
\author{P.Amore$^\dagger$, R.Cenni$^\ddagger$ and P.Saracco$^\ddagger$\cr
$\dagger$Dept.of Physics, College of William and Mary,\cr
 Williamsburg, VA 23185 USA \cr
$\ddagger$Istituto Nazionale di Fisica Nucleare  \cr
Dip.to di Fisica dell'Universit\`a,\cr
 Via Dodecaneso 33, 16146 Genova Italy}

\maketitle
\begin{abstract}
The presence of a resonant structure corresponding to the $\Delta$ 
excitation in the longitudinal response to an  e.m. probe is
investigated. It is shown that many-body effects could significantly
increase the relativistic contribution suggested by M.Ericson 
and coworkers
\end{abstract}

\section{Short introduction}

Few years ago Chanfray et al.\cite{ChDeErMo-93}  showed that the $\Delta 
(3,3)$ excitation is present also in
the longitudinal response of a nucleus to an electro-magnetic  (e.m.) 
probe.

The key issue in \cite{ChDeErMo-93} is elegant and simple. Let us assume 
that a nucleon at rest is excited to a $\Delta$ through a M1 
excitation. The above is purely transverse but since the 
longitudinal/transverse separation is frame-dependent, a
(relativistic) boost is able to give, in a moving frame, a longitudinal
component too. Since in a nucleus the nucleons are moving 
thanks to the Fermi motion, then the conversion
of a single (moving) nucleon to a $\Delta$ contains necessarily  a (small)
longitudinal component.

Here we wish to improve this analysis by accounting both for the 
possibility of creating a  $\Delta$ through a 
 E2 or C2 multipole  and (most relevant) for   
many-body effects, that can also lead to a (broadened) resonant structure.  

Concerning the former problem (developed in sect. 2) one has to face 
the limited experimental knowledge of the E2 and C2 excitations.  
The E2 current  is believed to be small (the coupling constant being 
expected, on the ground of the presently available data,
 to range between the -1\% and -2.5\% of the 
M1 one).
It is anyway negligible here, since
it is transverse too and of higher order than M1 in the non-relativistic 
reduction. Moreover the e.m. coupling constant is squared in the inclusive 
response so that a factor of about $10^{-4}$ is expected. 
Thus we shall not present in the following the effects of this contribution
since, as expected, our calculations 
 do not provide any perceptible 
change in the responses.

The C2 excitation is longitudinal and therefore could
play a role even
for small  coupling constants (in the Peccei 
lagrangian\cite{Pe-68,Pe-69}, for example, it
is about 10\% of the M1). It is thus accounted for in the following leaving
the corresponding coupling constant free to range between 0 and a 15\% of 
the M1 coupling constant.

The second problem, discussed in sect. 3, is of course by far more involved
and the calculations are consequently less firmly grounded. We make use in
the present paper of the Boson Loop Expansion (BLE) developed by two of us in 
some previous papers (see for instance \cite{CeCoSa-97}).

\section{The Relativistic Free Fermi Gas}

As it is well known, the longitudinal response is given by
\begin{equation}
R_L (q, \omega)= -{1\over \pi} {\rm Im} \Pi_{00} (q, \omega)
\end{equation}
where  
\begin{equation}
\Pi_{\mu \nu} (x, y) = {{<\psi_0| T \left\{j_\mu (x), j_\nu(y)\right\}| 
\psi_0>}
\over {<\psi_0 | \psi_0>}}\;
\label{eq:1}
\end{equation}
is the polarisation propagator (here written in configuration space).

The  $N$-$\Delta$ transition e.m. current operator, $j_\mu$, 
\begin{equation}
j_\mu^{M,E,C} (x) = f^{M,E,C}_{\gamma N\Delta}\overline{\psi}(x) {\cal 
O}_{\mu\nu}^{M,E,C}(-i\overleftarrow{\partial_x},-i\overrightarrow{\partial_x})
\psi^\nu_\Delta(x)\;,
\label{curr1}
\end{equation}
is expressed in terms of the nucleon and $\Delta$ fields, i.e.  
$\psi(x)$ and $\psi^\nu_\Delta(x)$ respectively.
The operators ${\cal O}^{M,E,C}_{\mu\nu}(-i\overleftarrow{\partial_x},
-i\overrightarrow{\partial_x})$
are usually written in Fourier transform, according to
\cite{JoSc-73,DeEiKo-76}, as
\begin{equation}
{\cal O}^M_{\mu\nu}(p,q) ={3(M_\Delta+M)\over 2 M M_\Delta Q^+}
\epsilon_{\mu\nu\lambda\rho}p^\lambda q^\rho
\label{m1m}
\end{equation}
\begin{equation}
{\cal O}^E_{\mu\nu}(p,q) =-{\cal O}^M_{\mu\nu}(p,q)
-{3\over 2M_\Delta^2
Q^+Q^-}{M_\Delta+M\over M} g_{\mu\rho}
\epsilon_{\nu\sigma\lambda\tau} p^\lambda q^\tau \epsilon^{\rho\sigma
\alpha\beta}p_\alpha q_\beta \gamma_5 
\label{e2m}
\end{equation}
\begin{equation}
{\cal O}^C_{\mu\nu}(p,q)  = -{3 \over 4 M_\Delta^2 
Q^+Q^-}{M_\Delta+M\over M}
(q^2 p_\mu-p\cdot q q_\mu)q_\nu \gamma_5\;;
\label{c2m}
\end{equation}
here $M_\Delta$ and $M$ are the $\Delta$ and nucleon masses respectively and
\begin{equation}
Q^\pm=\frac{q^2-(M_\Delta\pm M)^2}{2 M_\Delta}\;.
\label{qpm}
\end{equation}

In absence of interaction (Relativistic Free Fermi Gas model or simply RFFG) 
the evaluation of $\Pi_{\mu\nu}$ is rather simple:
\begin{eqnarray}
\lefteqn{\Pi_{\mu\nu}^{M,E,C}(q) =}\\
&&\nonumber - i \left(f^{M,E,C}_{\gamma N \Delta}\right)^2 
{\rm Tr}
\int {{d^4 k}\over {(2\pi)^4}} S_F (k) {\cal O}_{\mu\lambda}^{M,E,C}(q)
S^{\lambda\rho}_\Delta (k+q) {\cal O}_{\rho\nu}^{M,E,C}(-q)
\end{eqnarray}
$S_F$ being the usual in-medium relativistic propagator for 
nucleons, namely
\begin{equation}
S_F(k)={\rlap/{k} + M\over 2 E_k}\left\{{\theta(k-k_F)\over
k_0-E_k+i\eta}+{\theta(k_F-k)\over k_0-E_k-i\eta}
-{1\over k_0+E_k-i\eta}\right\}
\end{equation}
(where $E_k=\sqrt{k^2+M^2}$), and
$S^{\lambda\rho}_\Delta$ the usual Rarita-Schwinger propagator in the vacuum.

Symbolically $\Pi_{00}$ takes the form
\begin{equation}
\Pi_{00}^{M,E,C}=- i 
\int {{d^4 k}\over {(2\pi)^4}} {P^{M,E,C}(k,q)\over 
(k^2-M^2\pm i\eta)\left[(k+q)^2-M_\Delta^2+i\eta\right]}
\label{eq:2}
\end{equation}
where, using eqs.
(\ref{m1m},\ref{e2m},\ref{c2m}) and evaluating some complicated traces,
$P^{M,E,C}(k,q)$ turns out to be
\begin{equation}
P^{M,E,C}(k,q)=6\left(f^{M,E,C}_{\gamma N \Delta}\right)^2 
\frac{(M_\Delta+M)^2}{M_\Delta^2
M^2(Q^+)^2} \tilde P^{M,E,C}(k,q)
\label{pmec}
\end{equation}
with
\begin{eqnarray}
\lefteqn{\tilde P^M(k,q)=\left[{\bf q}^2 {\bf k}^2 -({\bf k}\cdot {\bf
q})^2\right] \left(M^2+M M_\Delta + q_0 E_k-{\bf k}\cdot {\bf
q}\right)
}
\label{M1}
\\
\lefteqn{\tilde P^E(k,q)=
\frac{1}{M_\Delta^2(Q^-)^2}
\left[{\bf k}^2 {\bf p}^2 - ( {\bf k} \cdot  {\bf q})^2\right]}
\label{E2}\\
&&\times\Bigl\{ - ( {\bf k} \cdot  {\bf q})^3
+( {\bf k} \cdot  {\bf q})^2 \left[M^2-M M_\Delta + 3 q_0 E_k\right]
\nonumber\\
&&+ 2 {\bf k} \cdot  {\bf q} q_0 M (M_\Delta-M) E_k
\nonumber\\
&&- {\bf k} \cdot  {\bf q} \left[M_\Delta^2 (Q^-)^2 + M^2 {\bf q}^2
+2 q_0^2 M^2 + 3 q_0^2 {\bf k}^2 \right]
\nonumber\\
&&+ q_0\left[M_\Delta^2 (Q^-)^2+ M^2 {\bf q}^2+q_0^2 {\bf k}^2
\right]E_k
\nonumber\\
&&+M\left[M_\Delta^2 (Q^-)^2(M+M_\Delta)+(M-M_\Delta)
\left(M^2 {\bf q}^2+q_0^2 {\bf k}^2\right)\right]\Bigr\}
\nonumber\\
\lefteqn{\tilde P^C(k,q)=
\frac{4}{ M_\Delta^4 (Q^-)^2} \left[q^2 M_\Delta^2-\left(q^2+q_0 
E_k-{\bf k}\cdot {\bf q}\right)^2\right]}
\label{C2}\\
\nonumber\\
&&\times \left[q_0 E_k-{\bf k}\cdot {\bf q}+M(M-M_\Delta)\right]
\left[q^2E_k-(q_0 E_k-{\bf k}\cdot {\bf q})q_0\right]^2
\nonumber
\end{eqnarray}

Even if eq.(\ref{eq:2}) appears to be divergent, it 
actually becomes finite when its value at $k_F = 0$ is subtracted
\cite{AlCeMoSa-88}. The removal of the divergences concerns in any case the 
real part only of the polarisation propagator. The longitudinal response at 
the end of the story (separating the three contribution M1, E2 and C2)
is obtaining by taking the imaginary part of the
$\Delta$-hole term of $\Pi_{00}$ and is the sum of the three terms, namely
\begin{equation}
  \label{rtemc}
  R_L^{(M,E,C)}=\int{d^3k\over (2\pi)^3}P^{M,E,C}(k,q)
\delta(k_0-E^\Delta_{q+k}+E_k)
\end{equation}
(each of them, of course, multiplied by its corresponding form factor)
where $E^\Delta_k=\sqrt{k^2+M_\Delta^2}$.

The M1 coupling constant has been chosen according to \cite{ChDi-89}, 
i.e. $f^{M}_{\gamma N \Delta}=2.72$, 
We have left instead the C2 coupling constant free to range between 0. and 0.4
(since no well established experimental data give serious indication about 
this
value, we ha chosen a rather high one: actually, as we shall see, it
does not significantly change the results, so that its uncertainty does not
really matter in the present case).

The free response functions for $^{12}C$, evaluated 
at $k_F=1.1 \ {\rm fm}^{-1}$ 
are displayed in fig. 1.
Here the Quasi-Elastic Peak (QEP) is treated relativistically and 
without dynamics (i.e., in RFFG
 approximation). 
 The $\Delta$ peak is evaluated 
according to the above equations.
\begin{figure}[htb]
\label{fig1}
\begin{center}
\mbox{
\begin{tabular}{cc}
\epsfig{file=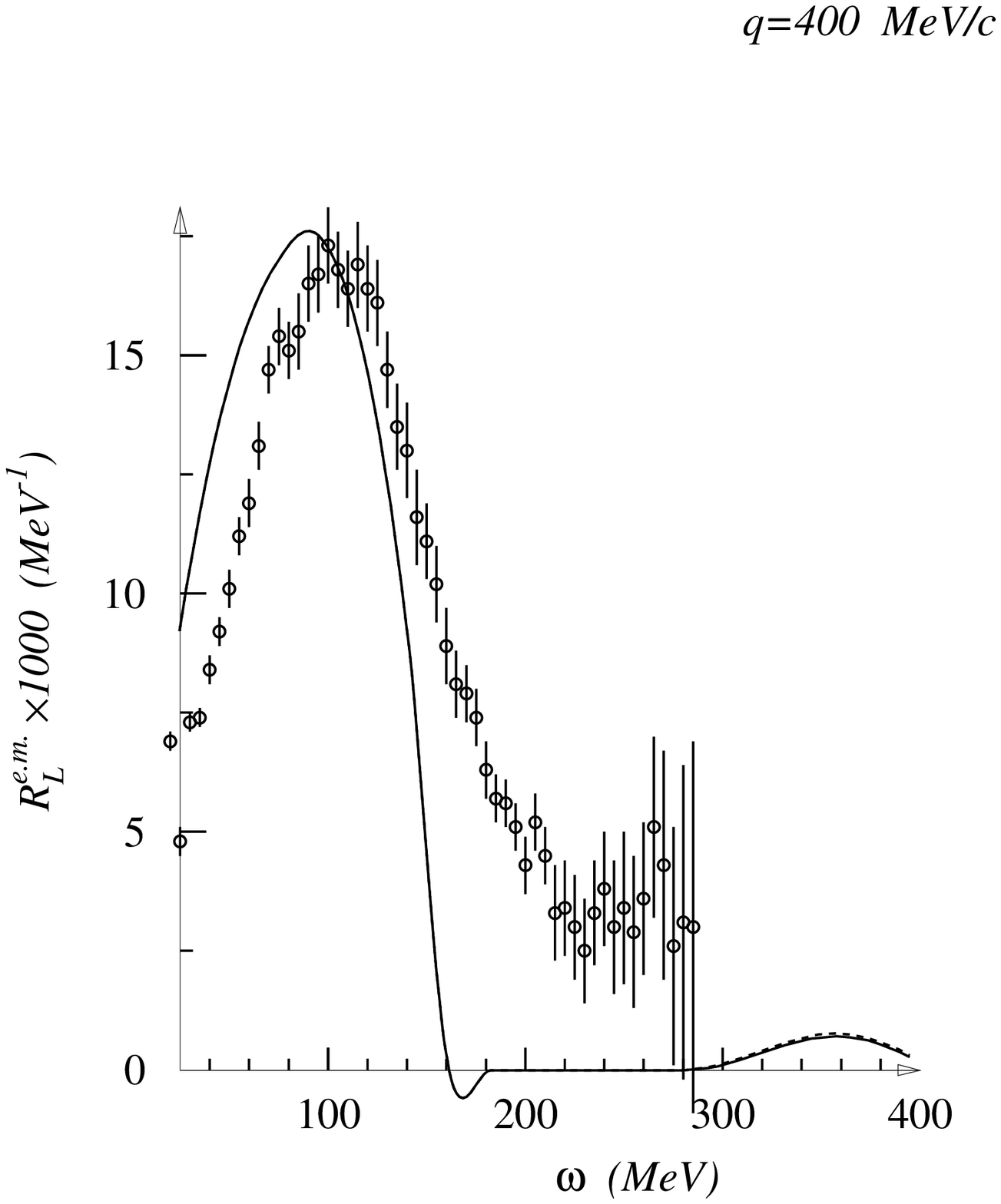,height=6cm}
&
\epsfig{file=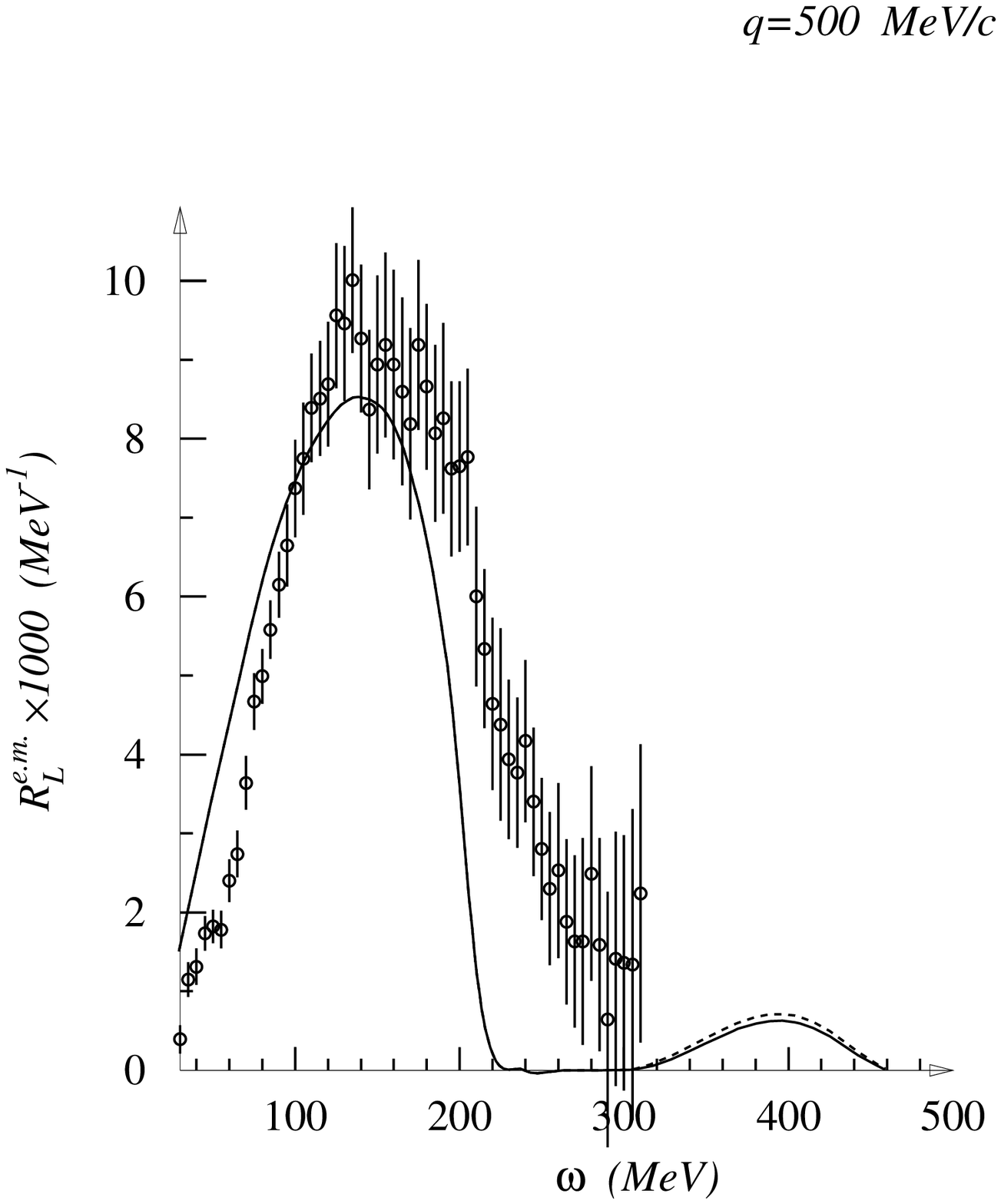,height=6cm}
\end{tabular}
}
\end{center}
\caption{The longitudinal response for an e.m. 
probe on $^{12}C$ at a transferred momentum $q=400$ MeV/c (left panel) 
and $q=500$ MeV/c (right panel): the solid line contains only the M1 
coupling while the dashed one 
(almost superimposed to the former) includes a C2 contribution with 
$f^C_{\gamma N\Delta}=0.4$.
The E2 contribution is not perceptible. 
Data from \protect\cite{Me-al-84,Me-al-85}}
\end{figure}

Since the e.m. transition form factors of the $\Delta$ 
are not so well determined as those of the nucleon \cite{St-93},
we have simply used the  standard nucleon dipole form factor for them.

In the figure the expected contribution appears as a small bump on the 
right of the QEP. The M1 contribution  here is by far 
dominant, the C2 term, even at its maximum coupling constant, being hardly 
distinguishable in the figure. 

\section{The contribution of the correlations}

We now come to deal with the  $\Delta$-excitations through 
many-body channels.
In the last two years, after a long time of 
unsatisfactory efforts, some calculations have been able to reproduce 
fairly well both the transverse and longitudinal responses in the 
QEP region (see 
\cite{CeCoSa-97,Fa-97,GiNiOs-97,AmCoLa-94} among others).
Here we adopt the approach developed by two of us, that naturally embeds 
the $\Delta$-degree of freedom into the formalism.

The theoretical frame has been already described in 
\cite{AlCeMoSa-87,CeSa-94,CeCoSa-97} and the
results for the e.m. response in the QEP are given in 
\cite{CeCoSa-97}.
We will therefore simply summarise our approach here.

The main theoretical tool in \cite{CeCoSa-97} is the 
bosonisation, which amounts to transform the original action
containing nucleons and $\Delta$'s as true degrees of freedom interacting 
via a potential into an effective bosonic one
(i.e. containing only bosonic degrees of freedom -- actually auxiliary 
fields introduced 
by means of a Hubbard-Stratonovitch transformation) fully equivalent 
to the former. Then the natural approximation scheme over the bosonic 
action is the 
boson loop expansion.
As described in \cite{CeCoSa-97}, the semi-classical limit (mean field) 
amounts to collect all diagrams having no boson loops and 
is  easily 
recognised to be the familiar  (not antisymmetrized) RPA.   
The next-to-leading order collects all the diagrams with
one boson loop: since fermion loops are of no matter in
determining the order of the BLE, then each boson line has to
be thought as an RPA-dressed potential. In this way a 
boson line embodies an infinity of p-h and $\Delta$-h insertions.
Thus not only a remarkable part of 2p-2h, but
also some 3p-3h and higher are included. However, not all the 2p-2h
contributions are accounted for: for instance, of the diagrams listed in ref.
\cite{AlDeDrMo-91} the IVb) and Vb) of fig. 9 are absent.
More generally all diagrams with two (RPA-dressed) 
interaction lines and only one fermion loop are lacking. 
Instead, our approach warrants
the preservation of gauge invariance and of general theorems and sum rules.

Concerning a comparison with other approaches, the only possible one (being 
expressed in terms of Feynman diagrams too) is with ref. \cite{GiNiOs-97}.
In our case both higher order corrections to $\Delta$ self-energy,  
pion-in-flight and Kroll-Rudermann term are absent
(and this because the complications and the large amount of work they require
to be embedded in the formalism), while instead the possibility of having
$\Delta$-excitations in different fermionic loops (in practice the so-called
Box Diagrams of the Bonn potentials \cite{MaHoEl-87}) are accounted for,  
that not only play 
a relevant role in the QEP, as is well known, but are also fundamental in 
providing the response in the $\Delta$-h region. Note that the absence of 
part of the MEC causes an
underestimate of the response in the dip region, as will be apparent later in 
fig. 4. In any case at present no separate calculations for the 
longitudinal and
transverse responses in the $\Delta$ region have been provided by Oset and
coworkers. Still a comparison between our approach and the one of ref. 
\cite{Fa-97} seems to be beyond our present possibilities.

Of central interest to us is that the 
 one loop order diagrams can contain $\Delta$'s even if the incoming 
and outgoing 
virtual photons interact only with nucleons. 
This originates a resonant structure, broadened
by a further p-h pair contemporarily present.  Typical diagrams  are
 displayed in fig. 2. 
\begin{figure}
\label{fig2}
\begin{center}
\mbox{
\epsfig{file=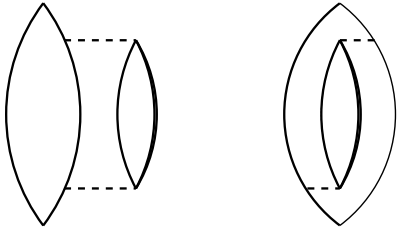,height=3cm}
}
\end{center}
\caption{Two typical many-body  $\Delta$-excitations 
in the charge-longitudinal channel. The thicker continuous line denotes
a $\Delta$ propagator, while in the present case the dashed line could 
represent a $\pi$ or $\rho$ exchange}
\end{figure}
There it is shown that a $\Delta$-h can be excited, with vector-isovector 
character, provided that another p-h pair 
again vector-isovector in nature re-establishes the correct 
quantum numbers.
Remarkably the whole class of diagrams considered in \cite{CeCoSa-97} and 
schematized in fig. 3 contains at least one $\Delta$-h pair, since
they surely appear in the RPA-dressed potential. 
\begin{figure}[htbp]
\label{fig:4}
  \begin{center}
    \leavevmode
    \epsfig{file=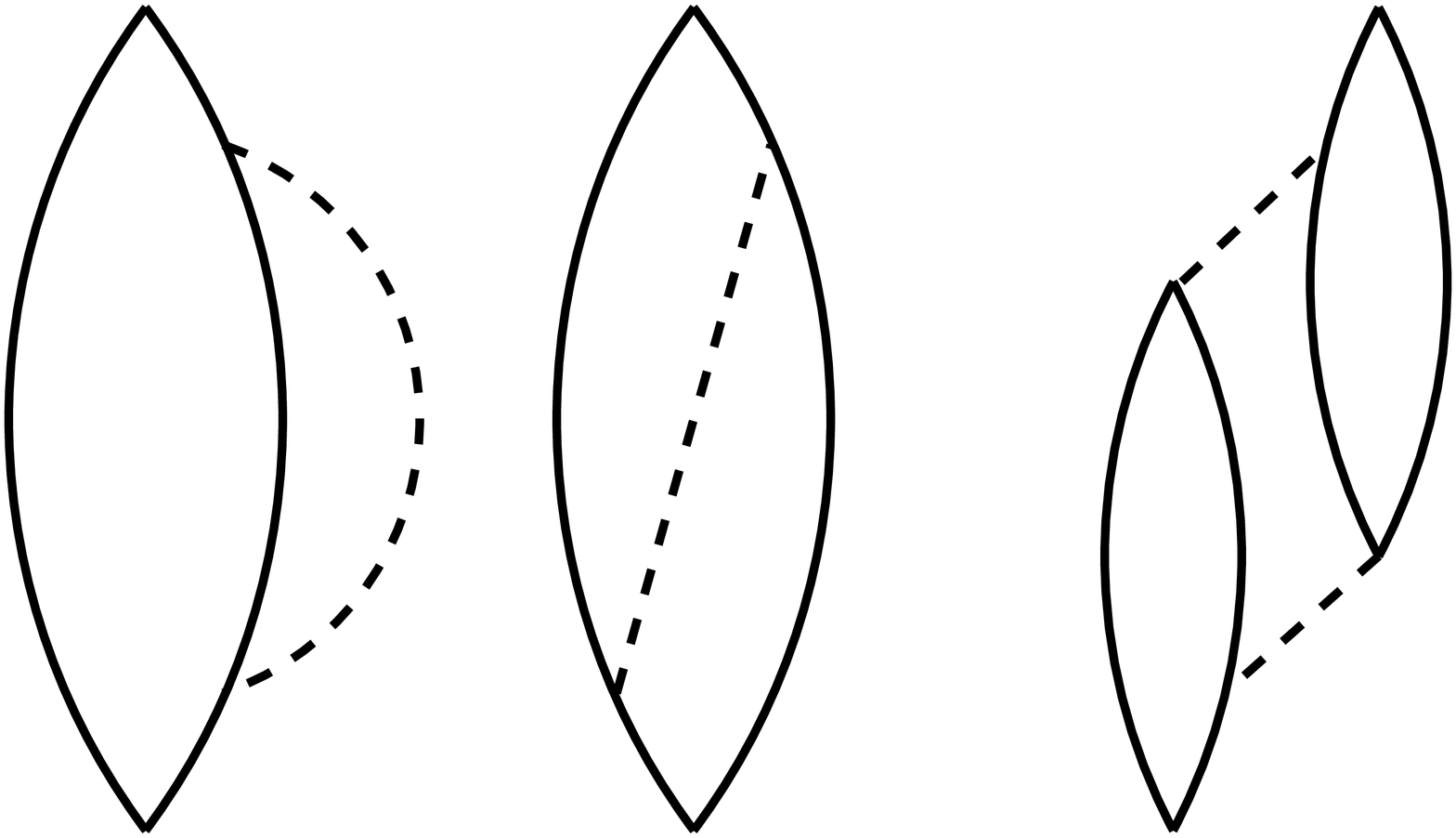,height=3cm}
    \caption{The classes of diagrams at the next-to 
leading order in BLE. Here the solid line could represent either a nucleon or 
a $\Delta$ while the dashed line this means an RPA-dressed meson propagator. }
  \end{center}
\end{figure}
There the dashed lines are the RPA-dressed, and consequently
also contain $\Delta$-h pairs , while the solid ones can be either a 
nucleon
or a $\Delta$ compatibly with the conservation laws. The vertices are 
assumed to be pure nucleon-nucleon or $\Delta$-$\Delta$ currents since this 
part of calculation is intrinsically non-relativistic.

The results of our calculations extend some partial outcomes hidden
in \cite{CeCoSa-97}: there the existence of a tail on the right of QEP was 
partially displayed but not discussed in one of the many figures, and, 
mainly, 
its resonant structure was not 
exhibited. Here the whole $\Delta$ peak is displayed and its  resonant
nature can be recognised -- an outcome not warranted {\em a priori} -- and
the balance with the direct M1 excitation established.
Furthermore the crucial physical point is that
the effect discovered in \cite{ChDeErMo-93} is expected to rise with the 
transferred momentum, as the quoted paper shows, while the many-body 
corrections  are expected to decrease.
\begin{figure}
\label{fig3}
\begin{center}
\mbox{
\begin{tabular}{ccc}
\epsfig{file=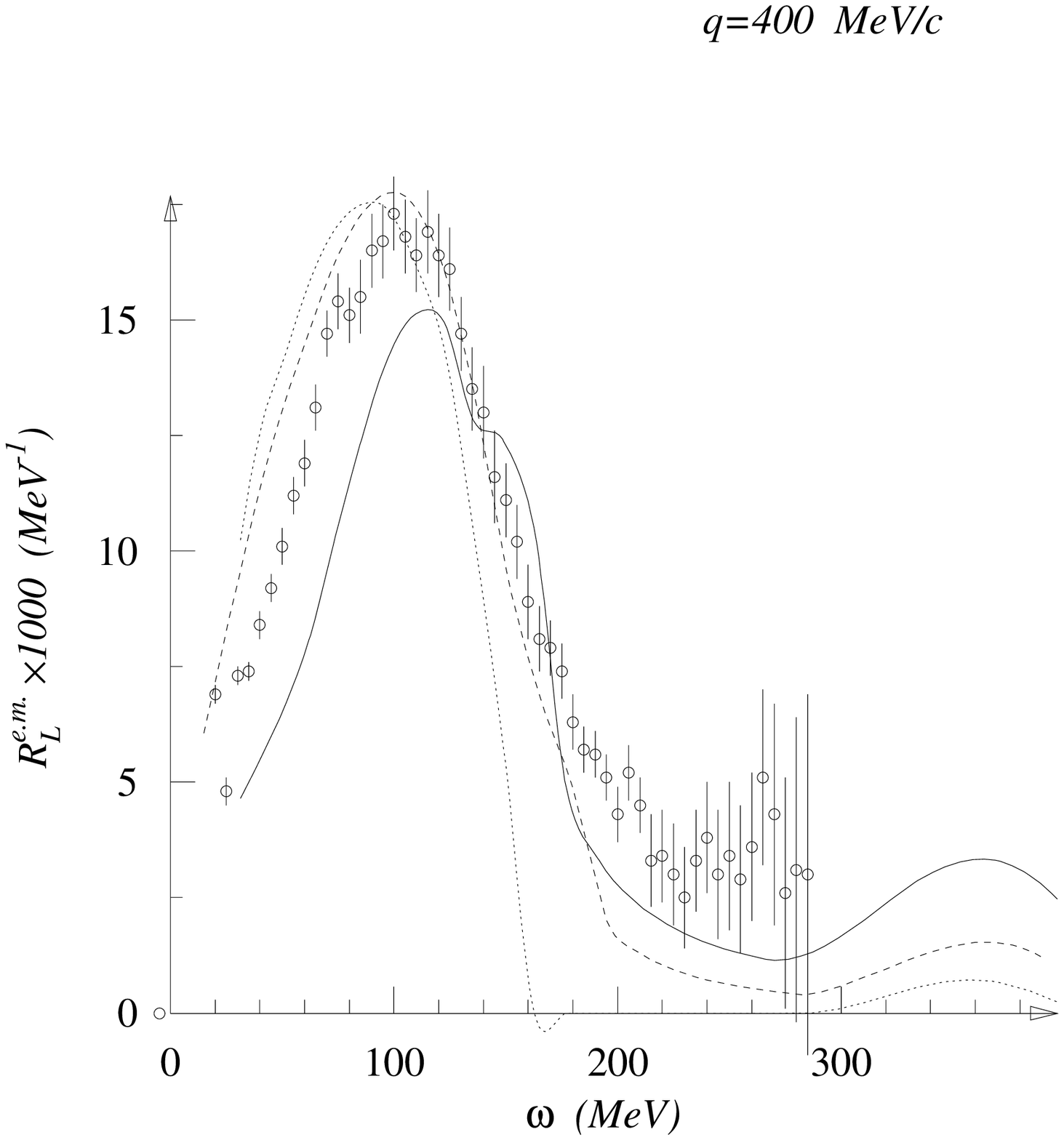,width=4cm,height=6cm}
&
\epsfig{file=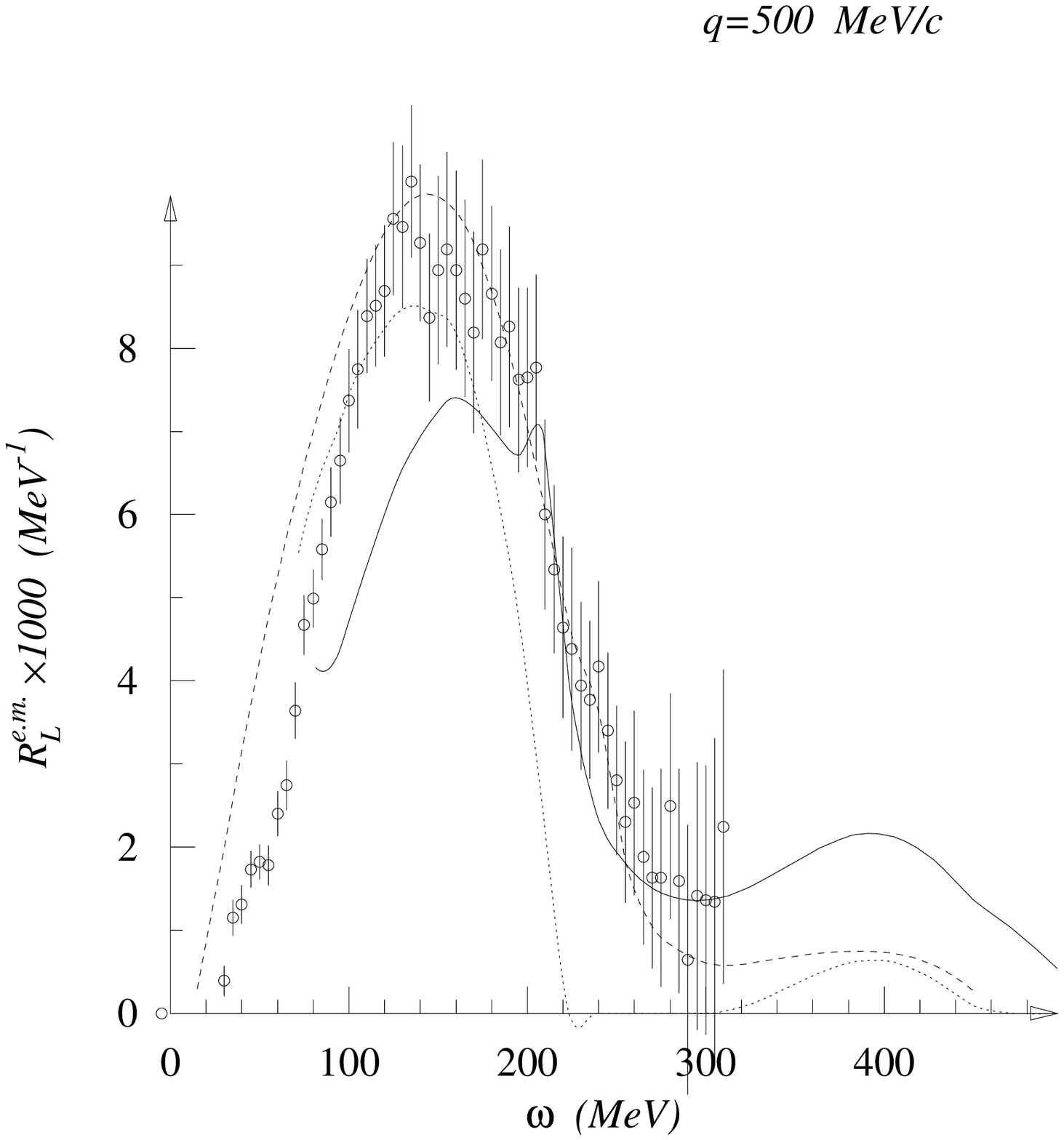,width=4cm,height=6cm}
&
\epsfig{file=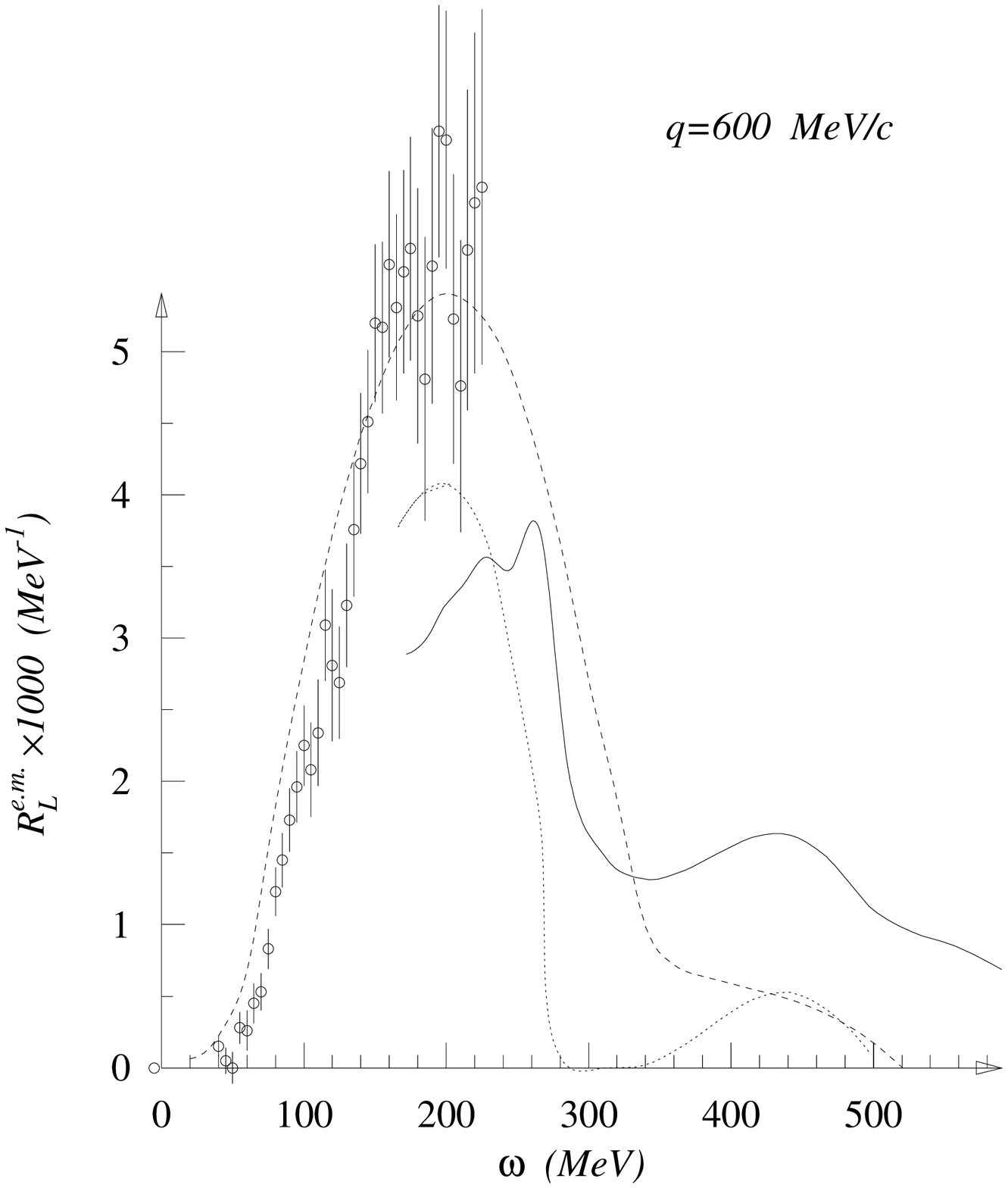,width=4cm,height=6cm}
\end{tabular}
}
\end{center}
\caption{The longitudinal response for an e.m. 
probe on $^{12}C$ at a transferred momentum $q=400$ MeV/c (left panel), 
 $q=500$ MeV/c (central panel) and $q=600$ MeV/c (right panel): 
the solid line describes  the full 
many-body calculation with a relativistic
mean field term plus the one-loop BLE calculation with the
kinematics adjusted as explained in the text (remember that the one-loop BLE
results are however intrinsically non-relativistic);
the dashed line refers to a fully non-relativistic calculation 
(hence the 0$^{\rm th}$ order contribution of the $\Delta$ is absent)
while the dotted line contains only RFFG plus free M1 response.
Data from \protect\cite{Me-al-84,Me-al-85}}
\end{figure}
The calculations are carried out with the same parameters given in 
\cite{CeCoSa-97}.
The results are shown in fig. 4.
Since the relativistic  kinematics can be of great relevance, because
 it tends to
shift both the QEP and the $\Delta$ one to lower energies, i.e., at greater
distances from the light cone, the electro-magnetic form factor will
also decrease somehow the height of the peaks. As repeatedly said 
throughout the
paper we are not able at present to perform a full relativistic calculation
of the one-loop BLE: the best thing we can do is to shrink somehow
the $x$-axis in fig. 4 in such a way to account for the relativistic 
kinematics, if not for the relativistic dynamics.
What we have done in practice is a rescaling of the $x$-axis
in such a way that the maximum of the QEP, the maximum of the $\Delta$
peak and the right corner of the free QEP will exactly coincide in energy
with the  position required by 
the relativistic kinematics. In particular it is extremely important
to fix the right corner of the free QEP because this strongly reduces the 
oscillations that are still observable in that point and that are due to the
failure of the BLE nearby a singular point. This procedure 
makes our pseudo-relativistic 
results in the left side of the QEP rather unreliable, but
on the other hand we are more interested, in the present paper, to the dip
and $\Delta$-peak region.

Let us now comment our main results, namely those displayed 
in fig. 4.
Here the solid line denotes a pseudo-relativistic calculation: this means
that the mean field is exactly evaluated in RFFG approximation both
in the QEP and in the $\Delta$-peak and that the one-loop correction are 
evaluated non-relativistically but the energy scale is shrinked as described 
above. The dashed line instead presents a fully non-relativistic calculation.
In this case coherently no free  $\Delta$-peak is accounted for and the 
corresponding bump on the right of the QEP comes solely from the one-loop 
corrections; finally the dotted lines describe the pure RFFG model.

As we expected from the very beginning the QEP is well reproduced by
the non-relativistic calculations, because the parameters of the model
(in practice the effective interactions in the various possible channels)
where adjusted on a non-relativistic dynamics.   Since relativity reduces
the free QEP the solid line underestimates it. On the other hand in
the $\Delta$-region the incoherence between the relativistic kinematics
and the non-relativistic treatment of the one-loop corrections is the 
price we must pay to
give a numerically acceptable description of the response.

Looking to this region, we see that the many-body excitations are 
clearly displayed and in general larger than the pure M1 response (again
a result not warranted {\em a priori} since both effects were expected to be
small). 
Another relevant question concerns the behaviour of the many-body correlations
at high momentum transfer. For this reason we have attempted to
perform a calculation
also for a transferred momentum of 600 MeV/c, where we know
that the non-relativistic
approximation is at least questionable. Disregarding in any case the QEP we see
(if in the $\Delta$-region our approach still can provide meaningful results)
that the fading out of the many-body correction is not so
strong as naively expected, and it opens the serious questions about how fast
the nuclear correlations will disappear. It suggests indeed that 
nucleon-nucleon
correlation will die much before than the $\Delta$-nucleon one.
Furthermore, the dashed line (that does not contain the free  $\Delta$-peak)
shows a clear resonant behaviour at a transferred momentum $q$= 400 MeV/c,
while the resonant structure seems to fully disappear at $q$= 600 MeV/c,
where instead the dip region is significantly filled.

Since these conclusions depend in a crucial way upon the behaviour of the 
response at high values of $q$, where our calculation suffer for the lack
of relativity, then the present
outcome clearly asks for fully relativistic many-body calculations, a task that
is at this moment forbidden to us due to some technicalities discussed in 
\cite{CeCoSa-97}.

Another relevant point is that even neglecting correlations
some strength is present in the resonant region 
and can give a significant (possibly of the order of some percent)
contribution to the Coulomb sum rule. This will  require  qualitative 
changes in the derivations of the former
to account for the $\Delta$ degrees of freedom as well.

Of course no serious comparison with the experimental data can be 
performed until a higher precision will be reached by the experimental data, 
such to enable the Rosenbluth separation even in a region 
where the longitudinal response is naturally depressed with respect of 
the transverse one. Nevertheless the indications coming from our 
calculation are surely not in contrast with the experimental data.

Needless to say new efforts, both on the theoretical and experimental 
side, are requested in order to understand this new -- and in our 
opinion interesting -- aspect of the nuclear dynamics.

\newpage

\end{document}